\documentclass[fleqn,usenatbib]{mnras}

\usepackage{newtxtext,newtxmath}

\usepackage[T1]{fontenc}
\usepackage{ae,aecompl}
\usepackage[normalem]{ulem}

\usepackage{graphicx}	
\usepackage{amsmath}	
\usepackage{amssymb}	

\usepackage{graphicx}

\usepackage[colorinlistoftodos]{todonotes}

\usepackage{upgreek}
\usepackage{array}
\usepackage{cite}
\usepackage{natbib}
\usepackage{float}
\usepackage{tabularx}
\usepackage{mwe}

\definecolor{mypink1}{rgb}{0.858, 0.188, 0.478}

\usepackage{microtype} 

\usepackage{fancyhdr}
\usepackage{mwe}
\usepackage[export]{adjustbox}[2011/08/13]

\defcitealias{Patricio}{P16}
\defcitealias{Floriane}{L17}

\def\OIIIUV{\mbox{O\,{\sc iii}}]~$\lambda \lambda 1661,66$}
\def\HeII{\mbox{He\,{\sc ii}}~$\lambda 1640$}
\def\CIV{\mbox{C\,{\sc iv}}~$\lambda \lambda 1548,51$}
\def\CIII{\mbox{C\,{\sc iii}}]~$\lambda \lambda 1907,09$}

\title[Spectral variations within Lyman-$\alpha$ haloes]{Spectral variations of Lyman-$\alpha$ emission within strongly lensed sources observed with MUSE}

\author[Claeyssens et al.]{
A. Claeyssens, $^{1}$\thanks{E-mail: \href{mailto:adelaide.claeyssens@univ-lyon1.fr}{adelaide.claeyssens@univ-lyon1.fr}}
J. Richard, $^{1}$
J. Blaizot, $^{1}$
T. Garel, $^{{1},{2}}$
F. Leclercq, $^{1,2}$
V. Patr\'icio, $^{3}$\newauthor{}
A. Verhamme, $^{2}$
L. Wisotzki, $^{4}$
R. Bacon, $^{1}$ 
D. Carton, $^{1}$
B. Cl\'ement, $^{1}$
E.C. Herenz, $^{5}$\newauthor{}
R.A. Marino, $^{6}$
S. Muzahid, $^{7}$
R. Saust, $^{4}$
J. Schaye, $^{7}$
\\
$^{1}$Univ Lyon, Univ Lyon1, Ens de Lyon, CNRS, Centre de Recherche Astrophysique de Lyon UMR5574, F-69230, Saint-Genis-Laval, France\\
$^{2}$Observatoire de Gen\`eve, Universit\'e de Gen\`eve, 51 Ch. des Maillettes, 1290, Versoix, Switzerland \\
$^{3}$DARK, Niels Bohr Institute, University of Copenhagen, Lyngbyvej 2, 2100 Copenhagen, Denmark\\
$^{4}$AIP, Leibniz-Institut f\"ur Astrophysik Potsdam (AIP) An der Sternwarte 16, D-14482 Potsdam, Germany\\
$^{5}$Department of Astronomy, Stockholm University, AlbaNova University Centre, 106 91 Stockholm, Sweden\\
$^{6}$ETH Zurich, Department of Physics, Wolfgang-Pauli-Str. 27, CH-8093 Zurich, Switzerland\\
$^{7}$Leiden Observatory, Leiden University, P.O. Box 9513, 2300
RA, Leiden, The Netherlands\\
}

\date{Accepted  2019 September 2. Received 2019 August 30; in original form 2019 May 28}

\pubyear{2019}

\begin{document}
\label{firstpage}
\pagerange{\pageref{firstpage}--\pageref{lastpage}}
\maketitle

\begin{abstract}
We present an analysis of ${\rm H_{I}}$ Lyman-$\alpha$ emission in deep VLT/MUSE observations of two highly magnified and extended galaxies at $z=3.5$ and $4.03$, including a newly discovered, almost complete Einstein ring.
While these Lyman-$\alpha$ haloes are intrinsically similar to the ones typically seen in other MUSE deep fields, the benefits of gravitational lensing allows us to construct exceptionally detailed maps of Lyman-$\alpha$ line properties at sub-kpc scales.
By combining all multiple images, we are able to observe complex structures in the Lyman-$\alpha$ emission and uncover small ($\sim 120$ km\, s$^{-1}$ in Lyman-$\alpha$ peak shift), but significant at  > 4\,$\sigma$, systematic variations in the shape of the Lyman-$\alpha$ line profile within each halo.
Indeed, we observe a global trend for the line peak shift to become redder at large radii, together with a strong correlation between the peak wavelength and line width.
This systematic \emph{intrahalo} variation is markedly similar to the object-to-object variations obtained  from the integrated properties of recent large samples.
Regions of high surface brightness correspond to relatively small line shifts, which could indicate that Lyman-$\alpha$ emission escapes preferentially  from regions where the line profile has been less severely affected by scattering of Lyman-$\alpha$ photons.
\end{abstract}




\section{Introduction}
\label{intro}
Galaxies are surrounded by a large amount of neutral hydrogen that forms part of the circumgalactic medium (hereafter CGM), the interface through which a galaxy interacts with its environment \citep{Tumlinson17}.
The physics of the CGM is key to explain how galaxies acquire gas and evolve.

The presence of CGM gas around high-redshift galaxies has been revealed through Lyman-$\alpha$ absorption seen in the spectra of background quasars \citep{Adelberger2005,Steidel2010,Rudie13,Turner14}. It is also detected through Lyman-$\alpha$ emission at several kpc scales, where photons scatter resonantly and illuminate the surrounding hydrogen gas, producing an extended Lyman-$\alpha$ halo  (hereafter LAH, \citealt{Steidel11,Herenz15,Lutz16,Momose16,Floriane}, hereafter L17). This goes even up to $100s$ of kpc for studies surrounding high redshift quasars such as the SLUG nebula \citep{Leibler18} or the COS haloes \citep{COShaloes}.

The Lyman-$\alpha$ signal is complex and several mechanisms could be responsible for its production: scattering in the neutral gas \citep{Verhamme12,Gronke16} cold streams feeding the CGM \citep{Furlanetto05,Dijkstra09,Henry15}, the presence of satellite galaxies surrounding the main source of emission, or a combination thereof. Models of Lyman-$\alpha$ emission in idealised configurations such as expanding shells produce a diversity of spatially integrated Lyman-$\alpha$ line profiles in general agreement with the global observed spectra (\citealt{Verhamme2008,Gronke2017}), but which do not reproduce the spatial extension of LAHs (\citealt{Patricio}, hereafter P16).
More detailed predictions of Lyman-alpha emission from numerical simulations of galaxies exist at very high redshift (e.g. \citealt{Laursen2009, Yajima2015, Behrens2019,Smith2019}). However these generally produce symmetric broad Lyman-$\alpha$ lines and agree with observations only when including IGM absorption  at $z>5$. Currently, the only work which discusses spectral variations within Lyman-$\alpha$ haloes \citep{Smith2019} does not show very clear trends and focus on large (>5 kpc) scales.

It is therefore important to obtain spatially and spectrally resolved observations of the Lyman-$\alpha$ line, in order to disentangle these mechanisms and thus better understand the link between the galaxies and their close environment (as done previously by \citealt{Swinbank15,Prescott15}, \citetalias{Floriane}, \citealt{Kusakabe18}). This is typically performed at low redshift for example in the LARS sample \citep{LARS14}. 

However, at high redshift, the mapping of Lyman-$\alpha$ emission around individual galaxies is very difficult due to the sensitivity and resolution limits of current observational facilities. It remains a challenge to observe LAHs around high-redshift individual galaxies with a spatial resolution sufficient to perform a precise analysis of Lyman-$\alpha$ line variations in the CGM (a few dozens spatial regions in the halo with sub-kpc scales). For example, only the most extended objects (> 5kpc) from \citet{Floriane} in the UDF can be resolved for such a study. 
 
One way to investigate the spatial variations of Lyman-$\alpha$ profiles is to use strong gravitational lensing.
Lensing conserves surface brightness \citep{Etherington1933} (hereafter SB) but creates multiple, enlarged and distorted images of background galaxies.
By leveraging the power of lensing with the unique efficiency (end-to-end transmission of the instrument) of the Multi Unit Spectroscopic Explorer (MUSE) integral field spectrograph on the Very Large Telescope (VLT) \citep{MUSE}, we can observe $z>3$ Lyman-$\alpha$ emitters (hereafter LAEs) lensed by galaxy clusters, with improved spatial resolution (\citealt{Smit17,Vanzella17}).
Among these galaxies, the most extended and magnified are sufficiently well-resolved to perform a precise analysis of the Lyman-$\alpha$ line variations in the halo (down to $\sim 0.5$ kpc scales in the source plane).
Unfortunately, highly magnified systems at high redshift are extremely rare; only a few lensed galaxies at $z>3$ feature Lyman-$\alpha$ emission subtending $>\!5$\arcsec{} on sky at a typical surface brightness limit of few $ 10^{-19}$ cgs (\citealt{Franx97}, \citealt{Smit17}, \citealt{Vanzella19}). Samples of highly magnified arcs are limited by the number of galaxies sufficiently extended intrinsically and lensed by a galaxy cluster.
So far, studies characterising the LAHs at high redshift have only reported minor variations in their spatial/spectral properties (e.g.\,\citealt{Erb18}). 

In this paper we present a detailed analysis of such spatial and spectral variations in two highly magnified LAHs:
a $z=3.5$ halo, previously presented by \citetalias{Patricio}, found in galaxy cluster SMACS\,J2031.8-2046 (hereafter SMACS2031) and another newly discovered $z=4.03$ halo behind the lensing cluster MACS\,J0940.9+0744 (hereafter MACS0940). All distances are physical.
We adopt a $\Lambda$ cold dark matter cosmology with $ \Omega_\Lambda = 0.7$, $\Omega_m = 0.3$ and $H_0=70$ km\,s$^{-1}\,{\rm Mpc}^{-1}$.

\section{Observations}
\label{obs}

The two selected Lyman-$\alpha$ emitters were known to be at $z>3.5$, highly magnified by galaxy clusters SMACS2031 \citep{Richard15} and MACS0940 \citep{Nicha}.
Observations for MACS0940 were performed as part of the MUSE guaranteed time observations between January 2017 and May 2018, with 33$\times$900-1000 seconds dithered exposures in WFM-NOAO-N (0.8 hrs) and WFM-AO-N (7.5 hrs) modes, for a total exposure time of 8.3 hrs. We covered a single 1$\times$1 arcmin$^2$ pointing sampled at 0.2\arcsec{} and centered on the cluster core. Conditions were photometric and the seeing was 0.60\arcsec{} at 700 nm as measured in the final datacube, which covers the wavelength range 475-930 nm with a spectral sampling of 1.25 \AA.
Observations of SMACS2031, obtained during 10hrs of MUSE commissioning, were previously presented in \citetalias{Patricio}.

Both datasets were (re-)reduced with the latest version of the MUSE data reduction software (\citealt{pipeline}, v2.4). We followed exactly the steps of the MUSE pipeline manual to perform basic calibration (such as bias, flat, wavelength, geometry) as well as science calibrations (flux and telluric correction, sky subtraction and astrometry). In particular we included the same self-calibration post-processing as the MUSE Ultra Deep Field (UDF, \citealt{UDF}), with some improvements to make it more robust on crowded fields like galaxy clusters. The idea of the self-calibration process is to correct for the IFU-to-IFU and slice-to-slice flux variations. It uses empty sky regions in the field to estimate flux correction per slice in several wavelength ranges and applies those correction factors after rejecting any outliers. This method can be used for galaxy clusters observations, as long as one provides a very clean mask of all objects detected in the field.
The final datacube was post-processed with the software {\sc zap} \citep{Soto} v2 to suppress the sky subtraction residuals, we used in this process the same object mask as for the self-calibration step.
These two additional treatments dramatically improved the commissioning data on SMACS2031 which were taken without any illumination calibration at the time, reducing the average variance measured in empty sky regions by 30\%.

Since the formal variances do not incorporate any covariance between adjacent pixels, these predicted variances are systematically too low, in consequence we rescale the variance cube. We followed the same method as \citet{Bacon15}: we selected a sample of random blank sky regions in the MUSE white image, where we measured the standard deviation within each region and between all these regions. We increased the MUSE variance by the square of the factor of these two measurements, scaled by the area in pixels of the empty regions considered. The effective variance is higher y a factor of 2.6 and 2.25 for SMACS2031 and MACS0940 respectively.

\section{Lens model}
\label{model}
We used the {\sc Lenstool} software \citep{Jullo07}\footnote{publicly available at \url{https://projets.lam.fr/projects/lenstool/wiki}} to perform a parametric model of the mass distribution in both cluster fields, where locations of strongly-lensed multiple images are used as constraints. The total mass distribution is parametrized as a combination of pseudo-isothermal mass profiles at cluster and galaxy scales (e.g. \citealt{2010MNRAS.404..325R}). The model of SMACS2031 is based on  \citet{Richard15}, with some improvements to the optimisation available in the latest version of {\sc Lenstool}.
The model of MACS0940 is constrained by two spectroscopically-confirmed multiple systems at $z=4.0$ and $z=5.7$ identified in the MUSE data and producing four images each. We describe each lens model with more details in Appendix \ref{annexe}. For the rest of this study we use the best model which minimizes the distance between the observed and predicted locations of multiple images (model rms of 0.33\arcsec{} and 0.23\arcsec{} for SMACS2031 and MACS0940 respectively). The lens model allows us to precisely raytrace spatial locations between the source plane and the image plane, and estimate the total magnifications and relative errors (Table \ref{table1}). 

\begin{figure*}
    \centering

    \begin{minipage}{0.45\textwidth}
    \includegraphics[width=\linewidth]{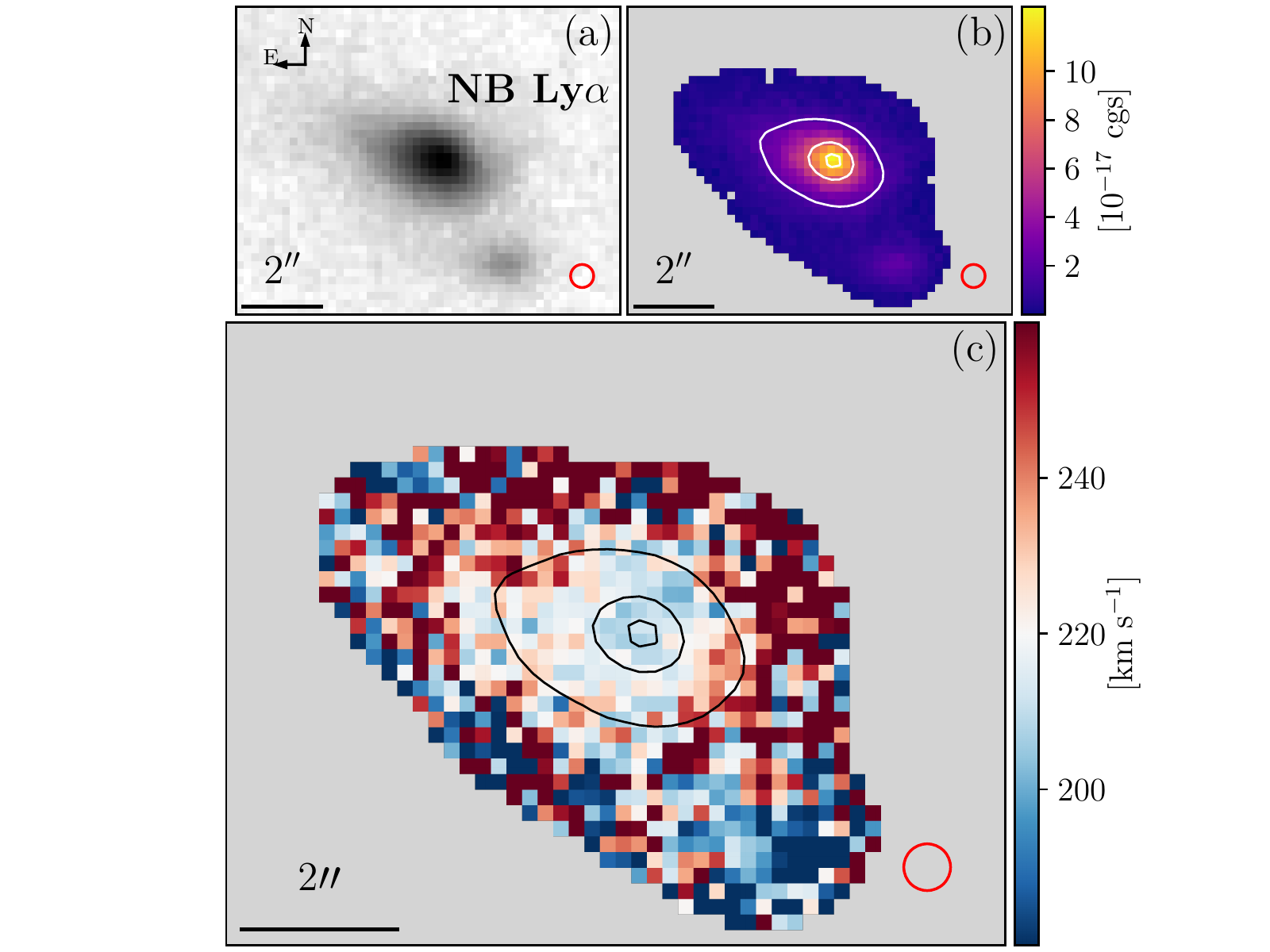}
    \end{minipage}
    \hspace{-0.7cm}
    \begin{minipage}{0.57\textwidth}
    \includegraphics[width=\linewidth]{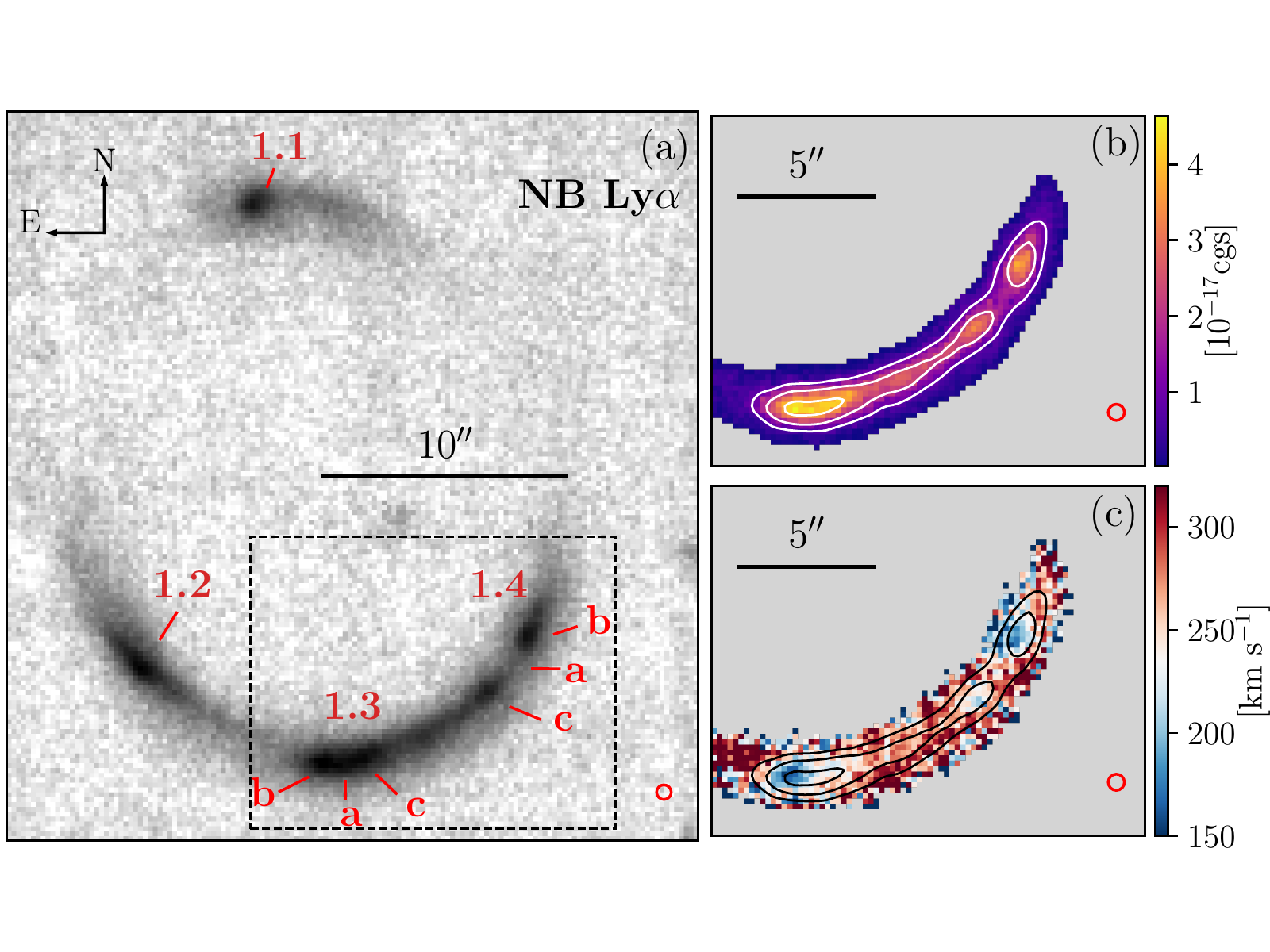}
    \end{minipage}
    \caption{Narrow-band Lyman-$\alpha$ image and pixel-by-pixel Lyman-$\alpha$ analysis in the image plane  for SMACS2031 (left panel) and MACS0940 (right panel). For each panel: (a) Narrow-band Lyman-$\alpha$ image of the entire arc (MACS0940, labels mark the multiple systems presented in the Table. \ref{mul_0940}}) and only one image of SMACS2031 (image 1.3 according to the notation used in \citealt{Richard15}), results of the individual spaxel fits (Lyman-$\alpha$ SB level (b) and peak shift (c)) for the most magnified images. The red circles show the MUSE PSF, and we overplot SB isocontours to highlight the flux peaks. The dashed grey box in panel (a) for MACS0940 represents the field of view of the two maps for  this object. 
    
    \label{fig:pixels}
\end{figure*}
\section{Spatial/spectral analysis}
\label{analyse}
We applied an identical procedure to analyse the MUSE datasets for both sources.
We produced pseudo-narrow band (hereafter NB) images of the Lyman-$\alpha$ emission (Fig.\,\ref{fig:pixels}) by summing the continuum-subtracted datacube over $\sim$ $15$ \AA\  centred on the line.
We extracted a spectrum optimising the MUSE continuum redwards of Lyman-$\alpha$ ($\sim$1350-1650 \AA\ restframe) and measured the systemic redshift based on nebular emission lines (\HeII{}, \OIIIUV{}, \CIV{} and \CIII{}).
Global properties of both galaxies are presented in Table \ref{table1}.
We estimate the exponential scale radius $r_h$ of the LAH following \citetalias{Floriane} to perform a morphological fit.
The fit is done in two steps: first the UV continuum is fit by a 2D elliptical exponential profile (based on the MUSE continuum image). Then the Lyman-$\alpha$ halo is fit by two elliptical exponential profiles, fixed at the same spatial position. The scale radius of one of them is fixed to the continuum one. The optimised parameters are both amplitudes, as well as the scale radius of the second component. We took into account the lensing effect and the MUSE PSF in this fit (see appendix \ref{annexe}). This 2D fit of the halo is idealised because it makes the assumption that each object is only composed of one exponential component for the continuum and one for the Lyman-$\alpha$ emission. But this type of fit allows us to compare our results with the LAEs found in the UDF \citep{Floriane} and gives us a good estimation of the mean size of the Lyman-$\alpha$ halo and the continuum in the source plane.

Thanks to their high magnification ($\mu\sim19-33$), the total observed Lyman-$\alpha$ fluxes reach $\sim10^{-15}$ erg\,s$^{-1}$\,cm$^{-2}$, more than ten times brighter than any halo identified in the MUSE UDF \citepalias{Floriane}.  

The SMACS2031 galaxy produces five multiple images (labeled 1.1 to 1.5, \citealt{Richard15}). Image 1.1 is close to the cluster centre and highly contaminated by stellar light, we exclude it for the rest of this study.
The MACS0940 galaxy produces four multiple images (labeled 1.1 to 1.4, Fig.\,\ref{fig:pixels}) forming a spectacular, almost-complete, Einstein ring of 10" radius in Lyman-$\alpha$ (Fig.\,\ref{fig:pixels}), covering $\sim 80$ arcsec$^2$ in the image plane.

\begin{table*}
    \begin{tabular}{cccccccccc}
                \hline  
                \hline
        Name & $z_{\rm sys}$ &$\mu_{\rm total}$& $r_h$&Ly$\alpha$ total flux    &Peak shift & FWHM& $a_{\rm{asym}}$ & $a$ & $b$ \\   
        &&&[kpc]&[10$^{-15}$ \,erg\,s$^{-1}$\,cm$^{-2}$]&[km$ \,$s$^{-1}$]&[km\,s$^{-1}$]&&&[km\,s$^{-1}$]\\
                \hline 
                SMACS2031	&$3.50618 \pm 0.00019 $&$	32.7 \pm2.8$& $1.5\pm 0.3$	&	$1.31 \pm 0.45 $ &$ 215 \pm 7$  & $274 \pm 6$ & $0.17 \pm 0.01 $ & $0.61 \pm 0.10$ & $52.0 \pm 28.1$\\
	MACS0940	&$4.03380 \pm 0.00056$ & 	$18.5\pm4.2$ & $4.3$ $\pm 0.2$ &	$1.16 \pm 0.54  $ & 	$240 \pm 7$& $441 \pm 8$ & $0.20 \pm 0.02$ & $0.80 \pm 0.06$ & $-108.8 \pm  28.6$\\
	         \hline  
     \end{tabular}
	 \caption{ General properties of the LAHs (on the total integrated spectrum). From left to right: redshift of the galaxy, total magnification (the magnification measurement is detailed in Appendix \ref{annexe}), halo scale radius (see text for details),  total observed Lyman-$\alpha$ flux (not corrected for magnification), best-fit parameters of the Lyman-$\alpha$ line (eq.\,\ref{eq1})  converted into km$ \, $s$^{-1} $ , slope and origin of the best fit by the \citet{Verhamme18} relation (${ \rm Peakshift = a \, FWHM  + b}$) presented in Fig.\, \ref{fig:plot_Anne} (see section \ref{analyse} for details)}. 
     \label{table1}          
\end{table*}
We then study Lyman-$\alpha$ line variations in the halo.
To model the Lyman-$\alpha$ line, we use the fitting formula:
    \begin{equation}
    f(\lambda) \ = \ {A}\,{\rm exp} \Big( -\frac{{(\lambda - \lambda_0)}^2}{2{(a_{{\rm asym}}\,(\lambda-\lambda_0)+d)}^2} \Big)
    \label{eq1}
    \end{equation}
introduced by \citet{Shibuya} to adequately model the asymmetric spectral profiles for LAEs. Using a simple gaussian symmetric fit instead would typically increase the final $\chi^2$ by $\sim 5$ in average. The free parameters of the fit are:  the line amplitude $A$, the asymmetry parameter $a_{\rm{asym}}$, the typical width $d$, and the peak wavelength of the line $\lambda_0$.

To account for the underlying continuum, we measure the mean flux level bluewards and redwards of the spectral line. We include their contributions as a ramp function between the two continuum levels covering 6 \AA\ around the central wavelength of the line. We also tested an Heaviside function but found that the ramp function provided a more robust continuum estimate. We checked that the parameters chosen for this ramp function does not affect the final result.
To obtain a robust fit of a given spectrum and its associated variance we used {\sc emcee} \citep{emcee} which utilises a Markov chain Monte Carlo sampler to maximise the Gaussian likelihood of the modeled spectrum from the set of parameters and eq.\,\ref{eq1}. 
We fit the individual MUSE spatial pixels (hereafter spaxels) in turn, using a semi-empirical Bayesian approach.
We place broad Gaussian priors on each of the four parameters, where the mean of the prior is that derived from a fit to the total Lyman-$\alpha$ spectrum (integrated over the entire halo). We chose a gaussian dispersion of $250$ and $400$ km\, s$^{-1}$ as prior respectively for $\lambda$ (peak shift) and $d$ (width) parameters and of $10 \%$ and $50 \%$ for $A$ (amplitude) and $a_{\rm asym}$ (asymmetry) parameters.

In Fig.\,\ref{fig:pixels} we show the results of the fitting for SMACS2031 (image 1.3) and MACS0940 (images 1.3 \& 1.4).
However, while some coherent structure is observed, the maps become noisy in the outskirts of the halo. 

To increase the signal-to-noise, we spatially bin and combine matched regions in the multiple images together.
To achieve this, we first use {\sc lenstool} to obtain a parametric model  of the source flux distribution that simultaneously accounts for all multiple images and the effect of the MUSE Point Spread Function (PSF).
We used the \citet{Voronoi} tessellation to obtain source plane regions of minimum total flux in the source plane. This method optimally preserves the maximum spatial resolution of two-dimensional data given a constraint on the minimum flux in each bin. The Lyman-$\alpha$ spectrum in each region is then constructed by coadding the MUSE spaxels which have more than 20 per cent overlap with the raytraced region, this value of 20 percent is optimised to ensure that there is no gap between two adjacent spatial regions in the image plane. In doing so the central, and smaller regions only receive contribution from the most amplified multiple images. We check and manually join adjacent spatial regions to ensure a minimum signal-to-noise ratio $>5$ in each defined bin. We verify that variations in signal-to-noise ratio do not introduce systematics in the measured parameters. We also ensure that all bins in the image plane are spatially more extended than the PSF FWHM along at least one direction. We end up with 130 and 123 source plane bins for SMACS2031 and MACS0940 respectively.

\begin{figure*}
    {\includegraphics[width=0.6\linewidth]{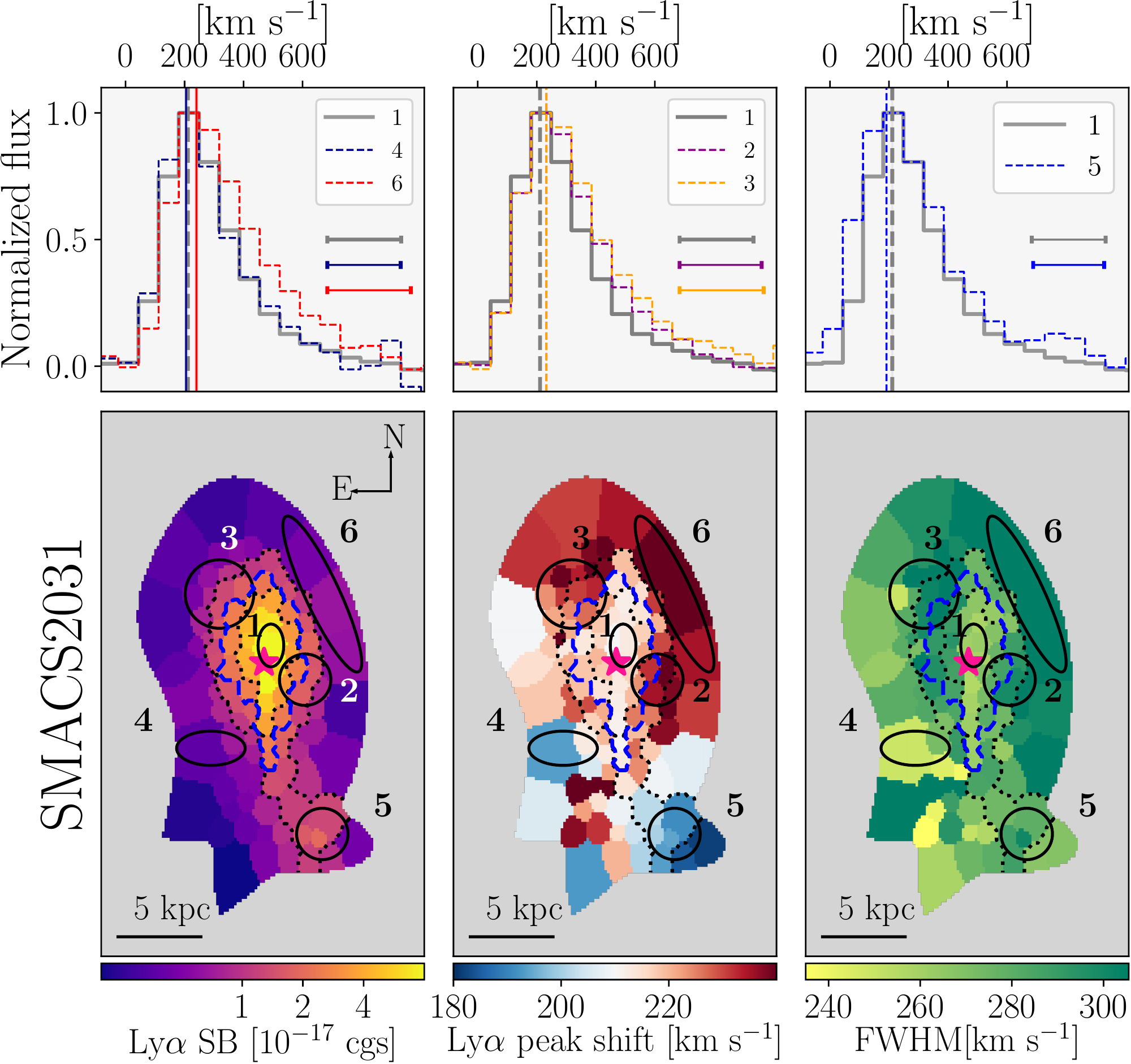}
    \includegraphics[width=0.8\linewidth]{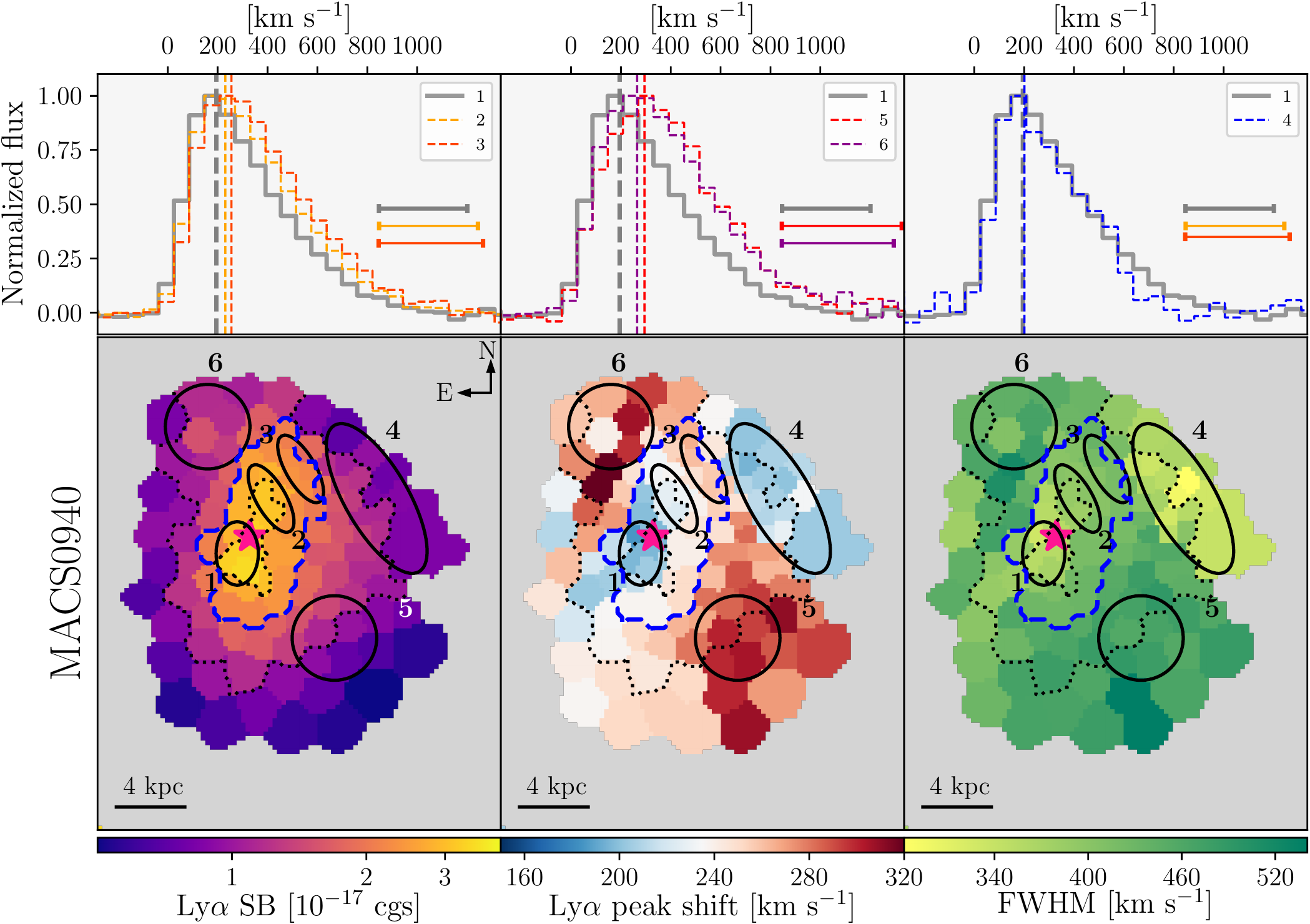}}
    \caption{
    Lyman-$\alpha$ spectral line analysis for SMACS2031 (upper panel) and MACS0940 (lower panel) in the source plane. For each object, top: spectra of specific regions of the halo (indicated by black ellipses on maps). The vertical lines represent the position of the peak wavelength produced by the asymmetric fit. Horizontal segments represent the FWHM of each line. 
    Bottom left: maps of the Lyman-$\alpha$ SB profile in the halo.
    Middle: map of the shift of the Lyman-$\alpha$ peak relative to the systemic redshift. Bottom right: map of the FWHM in the halo. 
    In all bottom panels: dashed lines represent SB isocontours at 1, 2 and 3 $\times 10^{-17} \, {\rm erg} \ {\rm s}^{-1} \ {\rm cm}^{-2} \ {\rm arcsec}^{-2}$. The blue contour corresponds to the SB threshold used in Fig.\,\ref{fig:plot_horn},  i.e. 2 $\times 10^{-17} \, {\rm erg} \ {\rm s}^{-1} \ {\rm cm}^{-2} \ {\rm arcsec}^{-2}$. The pink star marks the position of the peak of the stellar UV continuum. 
    }
    \label{fig:MACS}
\end{figure*}

To check that the results are not sensitive to the accuracy of the lens model, we apply this method on individual multiple images, with exactly the same tessellation in the source plane, and we recover the same trends for Lyman-$\alpha$ line variations (the same min and max values happen at the same locations and the overall variations are similar within 20 km$ \, $s$^{-1} $).
Our results are also robust against choosing adifferent prior distribution (uniform or Gaussian), or changing the tessellation to use larger spatial bins.
Finally, we visually inspect each spectrum and fit results to check the fit. We ensure that the reduced $\chi^2$ of the fit (measured over the spectral line) is $<1$ in the very l arge majority (i.e. 123/130 regions in SMACS2031 and 117/123 regions in MACS0940 have a $\chi^2 < 1$ and only 2 regions in SMACS2031 have a $\chi^2 > 2$) of the spectra. This shows that the Lyman-$\alpha$ well reproduced everywhere in the halo with a simple asymmetric profile (eq.\,\ref{eq1}) with no secondary line peak at bluer wavelengths. 
Figure \ref{fig:MACS} presents the resulting maps of Lyman-$\alpha$ peak shift and velocity dispersion in the source plane for both galaxies, where we convert $\lambda_0$ and $d$ from eq.\,\ref{eq1} into a velocity relative to the systemic redshift and FWHM respectively,  with the following analytic expression for FWHM: 
    \begin{equation}
    {\rm FWHM} \ = \frac{ \rm 2 \ \sqrt{2 \ \ln 2} \ d}{\rm 1 - 2 \ \ln 2 \ a_{\rm asym}^2}
    \label{eq2}
    \end{equation}

We also highlight the extracted spectra from specific regions to better illustrate the variations seen in the maps. 

\section{Results}
\label{results}

We have characterised the Lyman-$\alpha$ line properties in the haloes out to $10$\,kiloparsec (kpc) ($\sim\!2.1\,r_h$) in SMACS2031 and $10$\,kpc ($\sim\!2.5\,r_h$) in MACS0940.
SMACS2031 presents only mild variations of $\pm20\,$km\,s$^{-1} $ in peak shift and $\pm20$\,km\,s$^{-1}$ in the FWHM across the halo. MACS0940 presents stronger variations, with $\pm60$\,km\,s$^{-1} $ in peak shift and $\pm60$\,km\,s$^{-1} $ in FWHM. In these two objects we observe small, but nonetheless significant, variations in the Lyman-$\alpha$ line parameters at sub-kpc scales.
These variations are observed for the peak position ($\lambda_0$) and velocity dispersion $d$.
Indeed, on average, we can see that both $\lambda_0$ and $d$ increase towards large radii (i.e. the line gets redder and broader).
However, in both haloes, there are a few outer, low-SB regions that have relatively small peak shift ($\sim 200 $ km\,s$^{-1} $) comparable to the smallest value of the map. 

In SMACS2031, we identify two such regions, one of which (region 5 in Fig.\,\ref{fig:MACS}) has a distinct SB peak and as originally suggested by \citetalias{Patricio}, is probably a satellite galaxy.
In MACS0940 we observe a large region (no. 4) with a smaller velocity offset.
This could also potentially be a similar case of a companion, however, it does not show a local peak either in SB or in continuum.
When comparing both maps in Fig.\,\ref{fig:MACS} side by side, we can notice a strong link between Lyman-$\alpha$ peak shift and velocity dispersion. This correlation is evident when plotting one parameter against the other (Fig.\,\ref{fig:plot_Anne}). We also notice that high-SB regions have the smallest velocity offsets, while the opposite is not true (Fig.\,\ref{fig:plot_horn}).

\section{Summary and Discussion}
\label{conclu}

We used MUSE observations to analyse spectral properties of the Lyman-$\alpha$ line in two strongly lensed, extended LAEs at $z>3.5$. The emission line is always well-fit by a simple asymmetric spectral profile, redshifted from systemic.
The observation of a single red asymmetric peak in Lyman-$\alpha$ is generally assumed to arise from the presence of strong galactic winds ($>100$ km\,s$^{-1}$, e.g. \citealt{Verhamme06,Gronke16}).

We observe that the Lyman-$\alpha$ line profile is relatively consistent across the halo, the asymmetric Gaussian profile from \citep{Shibuya} reproduces very well the shape of the line with no secondary Lyman-$\alpha$ peak and with an almost constant asymmetry across the halo (0.15-0.25 in both cases). However we can observe significant variations of the other parameters at sub-kpc scales.
On average, at larger radii the peak shifts redwards and the line broadens.
Such spatial variations could be a result of the relative amount of H$_{\rm I}$ within the CGM, or its kinematics. 
We confirm for SMACS2031 the trends in Lyman-$\alpha$ variations across the halo already found by \citetalias{Patricio}, but with an improved spatial resolution when combining all images in the source plane. 
We acknowledge that the MUSE PSF introduces some correlation between adjacent source plane regions (Fig.\,\ref{fig:MACS}), but this is partially alleviated by combining multiple images at different shear orientations. 
Nevertheless, this means that the actual spatial variations seen in the Lyman-$\alpha$ line profile (in terms of peak shift and velocity dispersion) could be intrinsically stronger. 
    
We compare our results against resolved halo studies from the literature. \citet{Swinbank07} observed a similar object (a \mbox{$z=4.88$} galaxy lensed by the cluster RCS 0224-0002) with a single redshifted Lyman-$\alpha$ peak. They studied the source plane kinematics on 200 pc scales but did not find significant spatial variations of the Lyman-$\alpha$ peak shift across the halo. This was largely confirmed by \citet{Smit17} with MUSE/VLT observations.  However they only noticed minors variations of the Lyman-$\alpha$ line profile in a single outer region of the halo.

\citet{Erb18}, on the other hand, measured small variations of the Lyman-$\alpha$ line shape across the halo in a lensed galaxy at $z=2.3$. However, the double-peaked profile of its emission makes the comparison with our results complicated. More generally, object-by-object comparison is difficult and a larger sample would allow us to get a comprehensive view of the Ly$\alpha$ properties in the CGM.
    
\begin{figure}
    \centering

    \includegraphics[width=\linewidth]{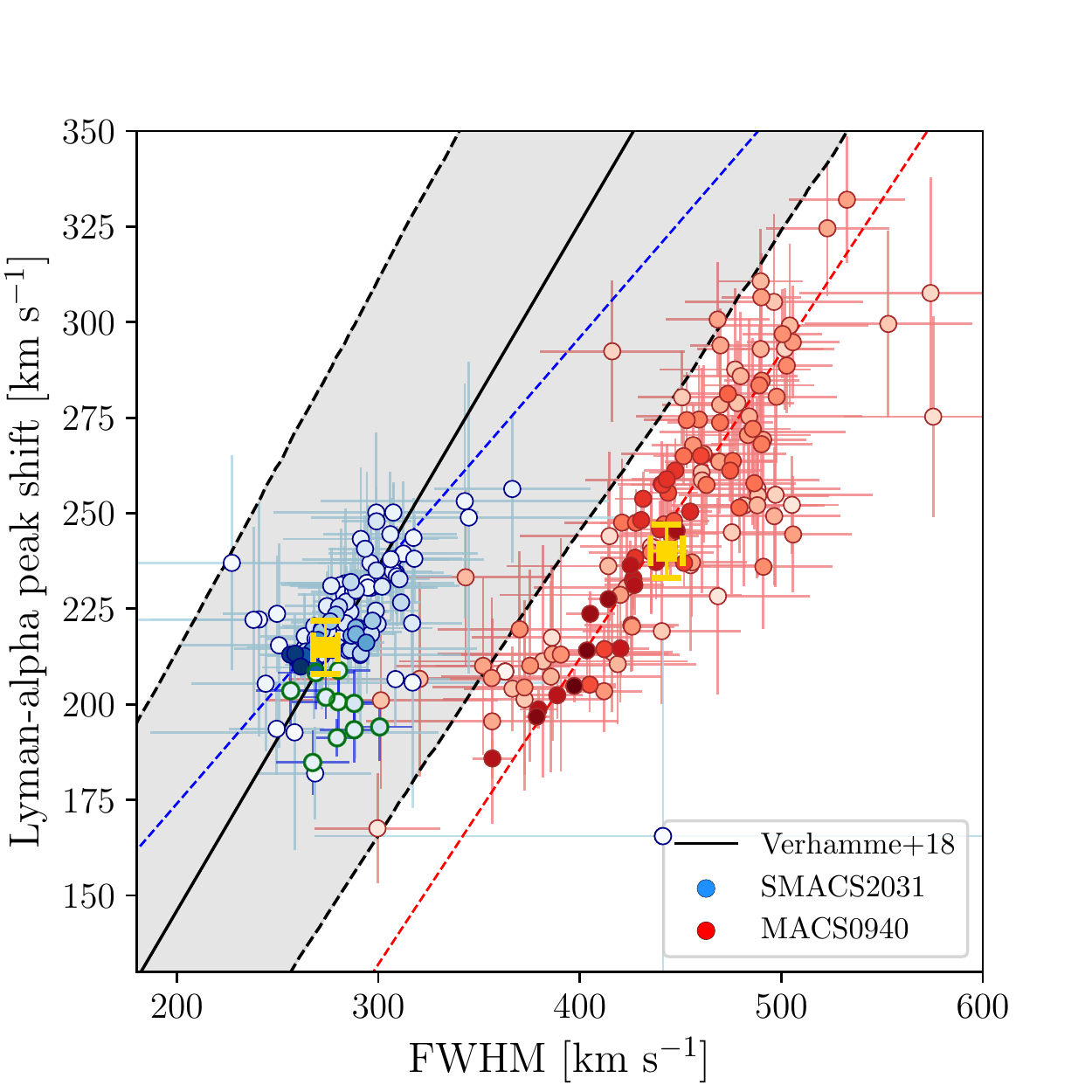}
    \caption{Lyman-$\alpha$ peak shift vs FWHM of the line in individual bins. Points are colour-coded in intensity according to the line SB. The green circles are regions from the companion in the SMACS2031 galaxy. Yellow crosses indicate the values  obtained for the fit to the total Lyman-$\alpha$ spectrum. The black solid line and shaded region represent the linear relation and errors found by \citet{Verhamme18}. The best fit parameters of this relation are presented in Table \ref{table1}.
    }
    \label{fig:plot_Anne}
\end{figure}

\begin{figure}
    \centering

    \includegraphics[width=\linewidth]{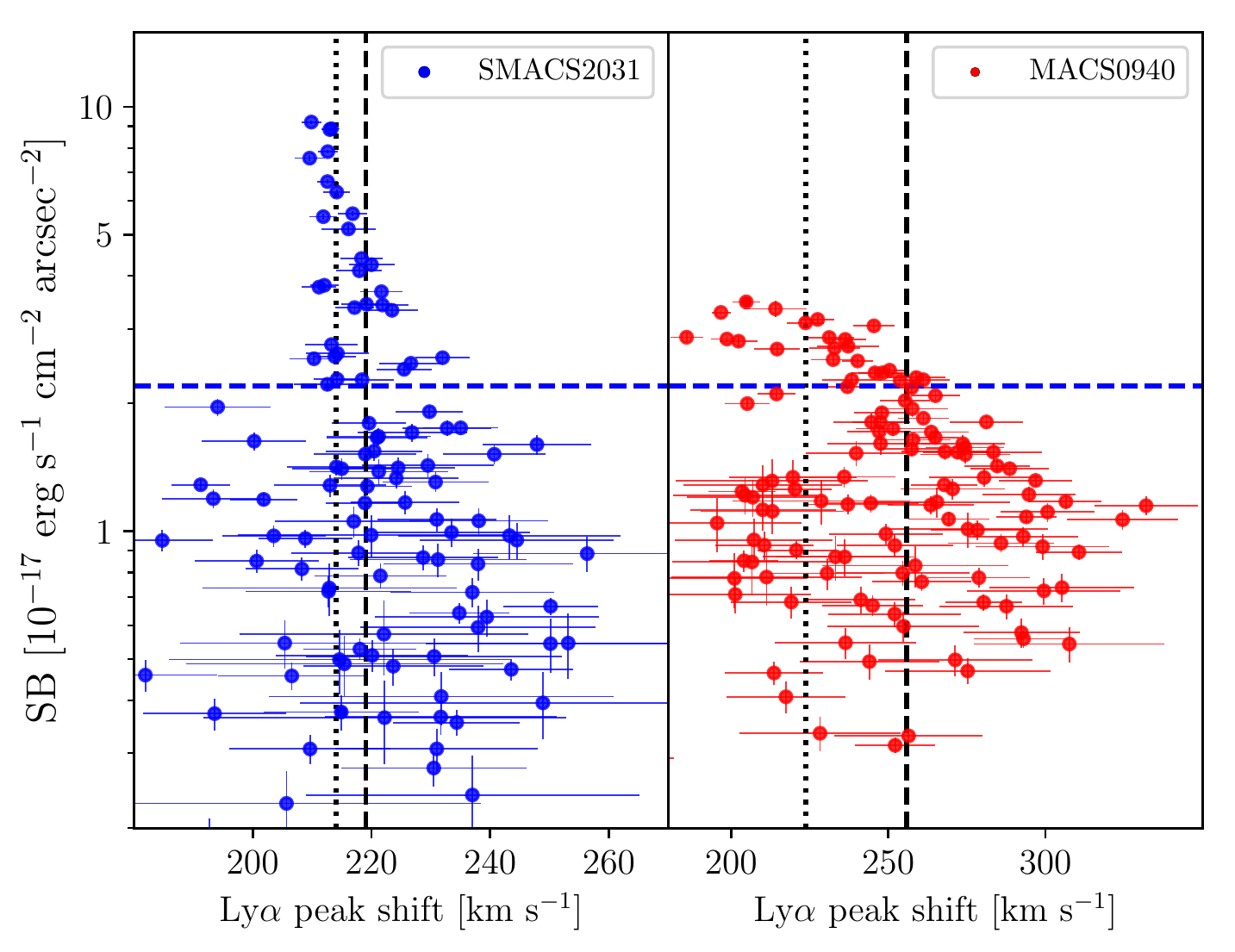}
    \caption{SB level as a function of Lyman-$\alpha$ peak shift for all regions in SMACS2031 and MACS0940. The green dashed line represents an arbitrary SB threshold at $2 \times 10^{-17} \, {\rm erg} \ {\rm s}^{-1} \ {\rm cm}^{-2} \ {\rm arcsec}^{-2}$ between high and low SB regions (same as the blue contour in Fig.\,\ref{fig:MACS}). The black dashed line highlights the mean peak shift at low SB and the dotted black line shows the mean peak shift velocity at high redshift. High SB regions are nearly always located at velocities smaller than low SB regions, significant at 4\,$\sigma$ (for SMACS2031) and 16\,$\sigma$ (for MACS0940)}. 
    
    \label{fig:plot_horn}
\end{figure}

In Fig.\,\ref{fig:plot_Anne} we overplot the empirical relation defined in \citet{Verhamme18} between Lyman-$\alpha$ peak shift and FWHM (not  corrected for the line spread function) obtained with large samples on an object-by-object basis. Due to the uncertainties in the systemic redshift, the values of peak shift could be biased by $\pm13$ km$ \, $s$^{-1} $ and $\pm33$ km$ \, $s$^{-1} $ for SMACS2031 and MACS0940, respectively; but this does not affect our results on variations within the halo and the slope of the correlation.. We can see that the correlation between peak shift and FWHM within each object follows the same empirical relation (in particular the same slope) as the one established on an object-by-object basis. This becomes even more visible when excluding the regions from the companion in SMACS2031. We measure Pearson correlation coefficients of $\rho=0.4$ for SMACS2031 (excluding the companion) and a value of $\rho=0.5$ for MACS0940 (p-value $<0.0001$ in both cases). We note that the MACS0940 regions are located below the empirical relation but very close to the 1\,$\sigma$ error so are marginally consistent. We find for both objects a linear slope (Table \ref{table1}) close to the \citet{Verhamme18} relation ($a=0.9$). The linear fit of the two datapoints series was also performed with the {\sc emcee} package accounting for measurement errors along both directions.
It is worth noting how similar the  slopes are for both sources, which suggests that the global offset could be due to a process linked with another galaxy parameter.
Lyman-$\alpha$ FWHM and peak shift are intrinsically linked due to radiative transfer effects within the CGM \citep{Verhamme06}, and here we show for the first time this effect within internal regions of LAEs as opposed to only from galaxy to galaxy. 
Figure \ref{fig:plot_horn} shows the spatially resolved relation between SB and peak shift for each region. We can clearly see that for brightest regions of the halo, the peak shift is systematically lower. We show in Fig.\,\ref{fig:plot_horn} the mean Lyman-$\alpha$ peak shift for high and low SB regions in both objects. The average variations of the peak shift across the halo is $5  \pm 1$ km\,s$^{-1} $ for SMACS2031 and $32\pm 2$km\,s$^{-1} $ for MACS0940, both  significant at more than 4\,$\sigma$. Almost no point populate the top right corner of the plot, showing that Lyman-$\alpha$ photons preferentially escape from low-peak shift regions. This is explained if Lyman-$\alpha$ photons escape more favourably from regions where the line profile is less altered, i.e. at small velocities or encountering a lower hydrogen column density integrated along the photon path. On the contrary, photons are much more scattered when escaping from outskirt regions. 
A scenario in which Lyman-$\alpha$ photons are scattered through a wind accelerating as a function of radius could explain the global redshift and broadening of the line at low SB.
The presence of several low peak shift regions at high radius / lower SB indicate a complex structure of the CGM around the galaxy. In one case we are able to match such a region with a companion satellite galaxy, which could be offset in velocity.

Additional deep MUSE observations of lensing clusters will allow us to enlarge the current sample of very extended LAEs for which the same analysis can be performed.
Although these results are based on two extremely bright sources, they are intrinsically typical in terms of size and brightness of the ones found in the UDF \citepalias{Floriane}. Observing spatial variations in such haloes has only been achievable so far using lensing magnification. The trends could, however, be tested on the brightest and most extended sources without lensing (Leclercq et al. in prep.).   

\section*{Acknowledgements}
We thank the anonymous referee for their helpful report.
AC, JR, DC and BC acknowledge support from the ERC starting grant 336736-CALENDS. TG and AV acknowledges support from the European Research Council under grant agreement ERC-stg-757258 (TRIPLE). VP is supported by the grant DFF - 4090-00079. This research made use of Astropy,\footnote{http://www.astropy.org} a community-developed core Python package for Astronomy \citep{astropy:2018}. Based on observations made with ESO Telescopes at the La Silla Paranal Observatory
under programme ID 060.A-9100, 098.A-0502, 0100.A-0249, 0101.A-0506.




\bibliographystyle{mnras}
\bibliography{sample}



\appendix

\section{Lens models and uncertainties}
\label{annexe}
We present here in more details the two lens models used in our analysis. The procedure described uses the Lenstool software and is similar to previous cluster mass models from our team \citep{Richard10,Jauzac16,Malher18}.
We used the sky positions and redshifts of multiple images to constrain a parametric mass model of each cluster. We adopt a dual pseudo-isothermal elliptical mass distribution (dPIE, \citealt{dPIE}) which is an isothermal profile  to model the different components of the mass model (dark matter halo and cluster galaxies). These dPIE components are parametrized by a position ($x$,$y$), an ellipticity $\epsilon$, a position angle ($\theta$), a velocity dispersion ($\sigma$), a core radius ($r_{core}$) and a cut radius ($r_{cut}$). For the large majority of the cluster galaxies, we fixed the parameters ($x$,$y$), $\epsilon$ and $\theta$ at the values measured from their light distribution \citep{Kneib96} and assume empirical scaling relations (\citealt{Faber-Jackson1976} and constant mass-to-light ratio) to relate their velocity dispersion and cut radius to their observed luminosity \citep{Jauzac16}.
The $\chi^2$ is minimised based on the $rms$ between the  observed and predicted positions of multiple images by the model.

The model of SMACS2031 is based on the previous one published in  \citet{Richard15} (including the same set of 12 multiply-imaged systems used as constraints) with the following improvements. We used the latest version of Lenstool (v7.1) which includes more robust tests on the convergence of the model to perform the optimisation. We also include an additional \textit{external shear} component to account for unknown environmental effects in the mass distribution surrounding the region of multiple images. The new parameters for this model are presented in Table \ref{model_2031}.

The lens model for the cluster MACS0940 is totally new with MUSE. We used 2 lensed galaxies to do the optimisation, positions and redshifts of the multiple images are presented in Table \ref{mul_0940}. The best-fitting parameters of the model are presented in Table \ref{model_0940}.

To reconstruct the light distribution of the 2 lensed galaxies in the source plane we use the function {\sc shapemodel} in Lenstool. To do that we associated in the source plane an elliptical S\'ersic profile with each Lyman-alpha primary or secondary peak and fitted the position, ellipticity, position angle, effective radius and S\'ersic index. This parametrisation takes in account the lensing effect and the MUSE PSF. It is only used here to delimit regions maps in the source plane and is not used anywhere else.

Errors on lens and source parameters (Tables \ref{model_2031}, \ref{model_0940} and \ref{table1}) are computed with {\sc Lenstool} with a MCMC sampling the posterior probability distributions. The main source of uncertainty in the source reconstruction shown in the maps Fig.\,\ref{fig:MACS} is an overall scaling by $\pm 5$--$20 \%$ following the error on $r_h$ (Table \ref{table1}). However errors on the lens model do not affect the values from spectral fitting and the region to region variations seen in the maps. Neither do they affect the results on the peak shift and FWHM (Fig.\,\ref{fig:plot_Anne} and \ref{fig:plot_horn}).

\begin{table*}
\centering
    \begin{tabular}{cccccccc}
                \hline  
                \hline
                \multicolumn{8}{l}{SMACS2031 Reference $\alpha= 307.971900$ $\delta=-40.625225$ rms=$0.33"$    } \\
                \hline
        Component &$\Delta \alpha$& $\Delta \delta$ &$\epsilon$ /$\gamma$ &$\theta$& $\sigma_0$& $r_{cut}$ & $r_{core}$ \\   
      &["]&["]&&[deg]&[km$ \, $s$^{-1} $]&[kpc]&[kpc]\\
                \hline 
          DM1 & $0.34_{-0.10}^{+0.09}$ & $-0.82^{+0.10}_{-0.11}$ & $0.397^{+0.021}_{-0.019}$ & 2.4$^{+1.8}_{-1.5}$& $638^{+11}_{-11}$ & [1000] & 34.2$^{+1.7}_{-2.3}$ \\
          DM2 & 63.6$^{+0.4}_{-0.6}$ & 24.8$^{+0.9}_{-0.9}$ & 0.60$^{0.05}_{-0.05}$ & 5.5$^{+1.6}_{-2.0}$ & 1144$^{+17}_{-17}$ & [1000] & 149$^{+4}_{-5}$ \\   
                BCG & [+0.07] & [-0.054] & [0.092] & [-0.4] & 227$^{+4}_{-4}$ & 151$^{+2}_{-6}$ & [0.28] \\
                External Shear & - & - & $0.09^{+0.01}_{0.01}$ & $6.5^{+4.2}_{-3.4}$ & - & - & - \\ 
                 L* galaxy & - & - & - & - & 154$^{+7}_{-8}$ & 11$^{+2}_{-1}$ & [0.15] \\
                \hline
     \end{tabular}
	 \caption[model_2031]{Best-fitting model parameters for the SMACS2031 cluster lens model with two dark matter components (DM1 and DM2), 1 optimized cluster galaxies (BCG), 1 external shear (Ext. Shear) and the scaling relation of cluster members (shown for an L* galaxy.). From left to right: centre location in arcsecond from the reference location provided in each cluster, ellipticity, position angle, central velocity dispersion, cut and core radii of each dPIE profiles. Values between square brackets have been kept fixed during the optimisation.}
     \label{model_2031}          
\end{table*}

\begin{table*}
\centering
    \begin{tabular}{cccccc}
                \hline  
                \hline
        ID & $\alpha$& $\delta$ &$z_{spec}$&$\mu$ & Origin\\   
                \hline 
                1.1a & 145.22452 & 7.744060 & 4.03 & $3.2 \pm 0.7$ & HST F606W \\
                1.2a & 145.22574& 7.738704 & 4.03 & $8.4 \pm 5.9$ & HST F606W \\
                1.3a & 145.22366 & 7.737692 & 4.03 & $11.7 \pm 3.8$ & HST F606W \\
                1.3b & 145.22370 & 7.737670 & 4.03 & $11.1 \pm 3.4$ & MUSE Ly$\alpha$\\
                1.3c & 145.22328 & 7.737719  & 4.03 & $11.3 \pm 3.6$ & MUSE Ly$\alpha$\\
                1.4a & 145.22149 & 7.738897 & 4.03 & $9.3 \pm 2.7$ & HST F606W \\
                1.4b & 145.22138 & 7.739108 & 4.03 & $10.8 \pm 4.2$ & MUSE Ly$\alpha$\\
                1.4c & 145.22179 & 7.738490 & 4.03 & $5.2 \pm 2.5$ & MUSE Ly$\alpha$\\
                2.1 & 145.22615 & 7.742765 & 5.7 & $10.8 \pm 2.2$ & MUSE Ly$\alpha$\\
                2.2 & 145.22446 & 7.736915 & 5.7 & $3.7 \pm 0.3$ & MUSE Ly$\alpha$\\
                2.3 & 145.22142 & 7.741314 & 5.7 & $2.8 \pm 0.3$ & MUSE Ly$\alpha$\\
                \hline
     \end{tabular}
	 \caption[mul_0940]{Multiple image systems used in the lens model of MACS0940. From left to right we give their ID, positions, spectroscopic redshifts, magnification and from which image we measured positions. The arc in MACS0940 at $z=4.03$ is composed of 4 multiple images labelled from 1.1 to 1.4. The two most magnified images 1.3 and 1.4 are divided in 3 components: the continuum measured on HST (a), and 2 Lyman-$\alpha$ peaks labelled (b) and (c) (illustrated in Fig. \ref{fig:pixels}). The magnifications of each multiple images were computed with Lenstool and correspond to the magnification at the centre of the image.}    
     \label{mul_0940}          
\end{table*}

\begin{table*}
\centering
    \begin{tabular}{cccccccc}
                \hline  
                \hline
                \multicolumn{8}{l}{MACS0940 Reference $\alpha= 145.223740$ $\delta=7.740363$ rms=$0.23 "$    } \\
                \hline
        Component &$\Delta \alpha$& $\Delta \delta$ &$\epsilon$ /$\gamma$ &$\theta$& $\sigma_0$& $r_{cut}$ & $r_{core}$ \\   
      &["]&["]&&[deg]&[km$ \, $s$^{-1} $]&[kpc]&[kpc]\\
                \hline 
          DM & $0.088_{-0.451}^{+0.617}$ & $1.423^{+0.420}_{-0.723}$ & $0.579^{+0.092}_{-0.220}$  &$21^{+4}_{-9}$&$507.6^{+60.3}_{-24.7}$& [1000] & [25] \\
                BCG & [-0.101] & [0.055] & $0.153^{+0.126}_{-0.167}$ & $-26^{+27}_{-2}$ & $500.0^{+15.6}_{-99.8}$ & [52.1] & [0.077] \\
         Gal1 & [-11.781] & [3.075] & $0.117^{+0.279}_{-0.100}$ & $41^{+4}_{-79}$ & $108.3^{+177.0}_{-9.2}$ & 18.0 & 0.025 \\
                Gal2 & [6.026] & [-5.792] & [0] & [0] & $122.8^{+8.6}_{-22.5}$ & [50] & - \\
                External Shear & - & - & $0.0228^{+0.0354}_{0.0056}$ & $65^{+95}_{-19}$ & - & - & - \\ 
                 L* galaxy & - & - & - & - & [158] & [45] & [0.15] \\
                \hline
     \end{tabular}
	 \caption[model_2031]{Same as Table \ref{model_2031} but for MACS0940.}  
     \label{model_0940}          
\end{table*}


\bsp	
\label{lastpage}
\end{document}